\documentclass[a4paper,onecolumn,11pt,accepted=2021-07-12]{quantumarticle}
\pdfoutput=1
\usepackage[utf8]{inputenc}
\usepackage[english]{babel}
\usepackage[T1]{fontenc}
\usepackage{amsmath}
\usepackage{hyperref}
\usepackage[numbers,sort&compress]{natbib}

\usepackage{tikz}
\usepackage{lipsum}

\usepackage{mathtools}
\usepackage{physics}
\usepackage{xcolor}
\usepackage{graphicx}
\usepackage[left=23mm,right=13mm,top=35mm,columnsep=15pt]{geometry} 
\usepackage{adjustbox}
\usepackage{placeins}
\usepackage[T1]{fontenc}
\usepackage{lipsum}
\usepackage{csquotes}
\usepackage{amssymb}

\usepackage{dsfont}

\DeclareMathOperator*{\argmax}{arg\,max}

\newtheorem{theorem}{Theorem}
\newtheorem{lemma}{Lemma}
\newtheorem{proposition}{Proposition}

\begin{document}

\title{Emergent classicality in general multipartite states and channels}

\author{Xiao-Liang Qi}
\affiliation{Department of Physics, Stanford University, Stanford, CA 94305-4060, USA}
\author{Daniel Ranard}
\affiliation{Department of Physics, Stanford University, Stanford, CA 94305-4060, USA}
\maketitle

\begin{abstract}
In a quantum measurement process, classical information about the measured system spreads throughout the environment.  Meanwhile, quantum information about the system becomes inaccessible to local observers.   Here we prove a result about quantum channels indicating that an aspect of this phenomenon is completely general.  We show that for any evolution of the system and environment, for everywhere in the environment excluding an $O(1)$-sized region we call the ``quantum Markov blanket,'' any locally accessible information about the system must be approximately classical, i.e.\ obtainable from some fixed measurement.   The result strengthens
the earlier result of Brand$\tilde{\textrm{a}}$o et al.\ (\textit{Nat.\ comm.\ 6:7908}) in which the excluded region was allowed to grow with total environment size.  It may also be seen as a new consequence of the principles of no-cloning or monogamy of entanglement.  Our proof offers a constructive optimization procedure for determining the ``quantum Markov blanket'' region, as well as the effective measurement induced by the evolution.  Alternatively, under channel-state duality, our result characterizes the marginals of multipartite states.
\end{abstract}

\section{ Introduction}

By the monogamous nature of entanglement, a single quantum system cannot be highly entangled with many others.  From a dynamical perspective, this monogamy constrains the spreading of information.  The no-cloning theorem provides a simple example of such a constraint; more generally, quantum information cannot be widely distributed with high fidelity.

Constraints on information spreading also shed light on the quantum-to-classical transition.  Many questions remain about precisely how and when classical behavior emerges from quantum many-body systems.  When a small system interacts with a large environment, the environment often acts as a measuring apparatus, decohering the system in some basis.  This paradigm is further elaborated by research programs on decoherence and ``quantum Darwinism,'' describing how certain observables of the system are ``selected'' by the environment \cite{ollivier2005environment, blume2006quantum, riedel2010quantum, zurek2014quantum, zwolak2014amplification, riedel2016objective}.

Brand\~ao et al.\ \cite{brandao2015generic} proved a powerful monogamy theorem constraining the spread of quantum information. 
In a sense elaborated in Section \ref{sec:discussion}, they show that some aspects of the decoherence process must exist for \textit{any} quantum channel. They consider general time-evolutions of a system $A$ initially uncorrelated with a large multipartite environment $B_1 \otimes ... \otimes B_n$.  Their result states that for a large fraction of environmental subsystems $B_i$, the only information about $A$ that is accessible on $B_i$ must be classical, i.e.\ it must be obtainable from a fixed measurement on $A$.  Crucially, they show that the relevant measurement on $A$ is independent of the subsystem $B_i$ of interest.  Thus the system $A$ must ``appear classical'' to an observer at $B_i$, in the sense that the only accessible information about $A$ is classical.  

However, the abovementioned result only constrains a large \textit{fraction} of environmental subsystems.  For a fixed error tolerance, the number of subsystems left unconstrained by the theorem increases arbitrarily with the total size of the environment. Intuitively, this growth seems to contradict the monogamy of entanglement, which suggests the fragment of the environment with non-classical information about $A$ must have bounded extent. In other words, monogamy suggests the results of \cite{brandao2015generic} can be greatly improved.

In this paper, we obtain this stronger constraint on quantum information spreading. Our Theorem \ref{thm:channels} shows that for large environments, for everywhere in the  environment excluding some $O(1)$-sized subsystem $Q$, the locally accessible information about $A$ must be approximately classical, i.e.\ obtainable from some fixed measurement on $A$. This result corroborates the above intuition from monogamy.  The statement is totally general, applicable to arbitrary quantum channels and quantum states.  We call the excluded region $Q$ the ``quantum Markov blanket,'' or simply the Markov blanket, following the terminology in classical statistics \cite{pearl1988probabilistic}.

The proof of our result may be framed constructively as an optimization procedure, allowing numerical demonstrations on small systems. The central idea of the proof is to imagine expanding a small region of the environment to gradually encompass the entire system. During this process, one learns gradually more about the input system $A$. Through a greedy algorithm, one calculates an optimized path of expansion that extracts the most information from $A$. By strong subadditivity, even an optimal path must reach some ``bottleneck'' such that further expanding the region does not yield additional information about $A$. Analyzing this bottleneck gives rise to the result.  The simple mathematical argument is presented in Section \ref{sec:proof}, along with the path-based interpretation.

We also provide a numerical example involving a small spin chain in Section \ref{sec:examples}. Based on the proof of Theorem \ref{thm:channels}, our numerical algorithm identifies the quantum Markov blanket and the effective measurement induced on a subsystem by the dynamics. 

\section{Review} \label{sec:duality}

We briefly review quantum channels, channel-state duality, and measure-and-prepare channels.  Readers familiar with this material may wish to skip to the results in Section \ref{sec:main_result}, but the discussion relating static constraints like monogamy to dynamical constraints like no-cloning may still be of interest.

Recall that quantum channels describe the most general time-evolution of a quantum system, including interactions with an environment.  We denote a general quantum channel $\Lambda$ from system $A$ to $B$ as a map $\Lambda : \mathcal{D}(A) \to \mathcal{D}(B)$, where $\mathcal{D}(X)$ generally denotes the space of density matrices on system $X$.  Such a map is called a channel whenever it is completely positive and trace-preserving.

\subsection{ Channel-state duality }

The channel-state duality allows one to associate every channel with an essentially unique state, called the Choi state.  The correspondence defines a dictionary that translates between ``dynamical'' properties of channels and ``static'' properties of states.

In particular, given any channel $\Lambda : \mathcal{D}(A) \to \mathcal{D}(B)$, we construct the Choi state $\rho_{A'B}^{\Lambda}$, where $A'$ is a reference system isomorphic to $A$. We define
\begin{align*}
\rho_{A'B}^{\Lambda} = \Lambda(|\Gamma\rangle \langle \Gamma|_{AA'})
\end{align*}
by acting $\Lambda$ on subsystem  $A$ of an input state $|\Gamma\rangle\langle \Gamma|_{AA'}$ maximally entangled between $A$ and $A'$.  Different choices of maximally entangled pure state $|\Gamma \rangle$ yield different Choi states, related by unitaries on $A'$.\footnote{Alternatively, to avoid a choice of basis, we can identify the auxiliary system $A'$ as the vector space dual to $A$, denoted $A^*$.  Then the channel-state duality amounts to the observation that both (i) operators on $A^* \otimes B$ and (ii) linear maps from operators on $A$ to operators on $B$ may be interpreted as elements of $A \otimes A^* \otimes B \otimes B^*$. The less trivial aspect of Choi's theorem (see below) is then to identify the positivity of states with the complete positivity of channels.}

From the Choi state, we can recover the action of the channel as follows.  It is helpful to first choose bases; let $|\Gamma\rangle_{AA'}$ be the maximally entangled state
\begin{align*}
|\Gamma\rangle_{AA'} = \frac{1}{\sqrt{d_A}} \sum_i |i\rangle_A |i\rangle_{A'}
\end{align*}
with respect to some orthonormal bases $|i\rangle_A, |i\rangle_{A'}$. For any $\tau_A \in \mathcal{D}(A)$, define  $\tau_{A'} \in \mathcal{D}(A')$ so that $\tau_A$ and $\tau_{A'}$ are given by the same matrix in the $|i\rangle_A$ and $|i\rangle_{A'}$ bases, respectively.  Then we can recover the channel from the Choi state using the formula
\begin{align*}
\Lambda(\tau_A) = d_A \Tr_{A'}(\rho_{A'B}^{\Lambda} \tau_{A'}^T)
\end{align*}
where the transpose is taken in the $|i\rangle_{A'}$ basis.

Choi's theorem states that a linear map $\Lambda : \mathcal{D}(A) \to \mathcal{D}(B)$ is a channel iff the corresponding Choi operator $\rho_{A'B}^\Lambda$ is a quantum state with $\Tr_{B}(\rho_{A'B}^\Lambda)$ maximally mixed.  This correspondence is also called the Choi-Jamiolkowski isomorphism; see \cite{wilde2013quantum} for an extensive elaboration.

The channel-state duality allows one to relate dynamical and static properties.  The dynamical properties of a channel, characterizing information transfer from input to output, become static properties of the Choi state, characterizing correlations between the input (or rather the reference system) and the output.  

Constraints on dynamical properties of channels therefore entail constraints on correlation properties of states, and vice versa.  The equivalence of no-cloning and monogamy of entanglement provide a simple example. Because our main results constitute a more elaborate example, we explain this simple example first.

Consider a hypothetical cloning channel $\Lambda : \mathcal{D}(A) \to \mathcal{D}(B_1 \otimes B_2)$ with reduced channels $\Lambda_{B_1} : \mathcal{D}(A) \to \mathcal{D}(B_1)$ and $\Lambda_{B_2} : \mathcal{D}(A) \to \mathcal{D}(B_2)$ defined by $\Lambda_{B_1}= \Tr_{B_2} \circ \Lambda$ and $\Lambda_{B_2}= \Tr_{B_1} \circ \Lambda$. For $\Lambda$ to properly clone, we demand that $\Lambda_{B_1}$ and $\Lambda_{B_2}$ are identity channels.  However, under channel-state duality, reduced channels correspond to reduced states, and identity channels correspond to maximally entangled states.  So the Choi state $\rho_{A'B_1 B_2}$ must have $A'$ maximally entangled with both $B_1$ and $B_2$.  Hence the the no-cloning theorem (forbidding perfect cloning) automatically implies a simple monogamy theorem (forbidding maximal entanglement with two different systems), and vice versa.\footnote{This equivalence of results about no-cloning and monogamy also yields a simple operational picture, seen by unpacking the definition of the Choi state: if you could clone a system, you could violate monogamy of entanglement by cloning one half of a Bell pair.  The converse implication is slightly more involved: If you had a system $A$ maximally entangled with both $B_1$ and $B_2$, you could clone a system $A'$ by simultaneously teleporting it to both $B_1$ and $B_2$, by using the ordinary teleportation protocol but making use of both entangled pairs $\rho_{AB_1}$ and $\rho_{AB_2}$ simultaneously.}

\subsection{Measure-and-prepare channels}

An important type of channel for the subsequent discussion is the ``measure-and-prepare'' channel.  Such a channel takes the form
\begin{align} \label{eq:measure_and_prepare}
\rho \mapsto \sum_\alpha \Tr(M_\alpha \rho)  \sigma_\alpha
\end{align}
for some states $\{\sigma_\alpha\}$ and for some operators $\{M_\alpha\}$ that form a positive operator-valued measure (POVM), i.e.\ $M_\alpha > 0$ and $\sum_\alpha M_\alpha = \mathds{1}$.   Such a channel has the physical interpretation of performing a generalized measurement with some POVM $\{M_\alpha\}$ and then preparing a state $\sigma_\alpha$ determined by the measurement outcome $\alpha$.  Note the states $\sigma_\alpha$ are not required to be orthogonal, and they may even be identical, in which case the channel is constant and transmits no information about the hypothetical measurement outcome.

An important special case of measure-and-prepare channels is a ``quantum-classical'' channel.  Such a channel takes the form
\begin{align} \label{eq:QC}
\rho \mapsto \sum_\alpha \Tr(M_\alpha \rho) |\alpha\rangle \langle \alpha|
\end{align}
for some POVM $\{M_\alpha\}$ and orthonormal basis $|\alpha\rangle$.  (In what follows, a ``quantum-classical channel on system $A$'' refers to such a channel from $A$ to some auxiliary system spanned by $|\alpha\rangle$.) Likewise, a ``classical-quantum'' channel takes the form $\rho \mapsto \sum_\alpha \Tr(\rho|\alpha\rangle\langle\alpha|)\sigma_\alpha$.  A measure-and-prepare channel may then be seen as a quantum-classical channel (the ``measurement'') followed by a classical-quantum channel (the ``preparation''). 

A channel is measure-and-prepare iff it is ``entanglement-breaking,'' i.e.\ if it produces a separable state whenever it acts on one half of an entangled pair.  Relatedly, a channel is measure-and-prepare iff the Choi state is separable \cite{horodecki2003entanglement, korbicz2012quantum}.  For measure-and-prepare channels expressed as above in Eqn.\ \ref{eq:QC}, the Choi state takes the form (up to change of basis on the reference system)
\begin{align} \label{eq:measure_prepare_Choi}
\sum_\alpha p_\alpha \rho_\alpha \otimes \sigma_\alpha
\end{align}
where
\begin{align}
\rho_\alpha & = \frac{M_\alpha^T}{\Tr(M_\alpha)}, \nonumber \\
 p_\alpha & = \frac{1}{d_A} \Tr(M_\alpha). \nonumber
\end{align}
The expression is arranged so that the coefficients $p_\alpha$ form a probability distribution and the operators $\rho_\alpha$ are normalized states.  

We say two measure-and-prepare channels can be written using the same measurement if they use the same POVM $\{M_\alpha\}$.  Likewise, we say two separable states $\rho_{AB_1}$ and $\rho_{AB_2}$ can be written using the same ensemble of states $\{p_\alpha,\rho_{\alpha}^A\}$ on $A$ if they both take the form 
\begin{align} \label{eq:same_ensemble}
\rho_{AB_1} & = \sum_\alpha p_\alpha \rho^A_\alpha \otimes \sigma^{B_1}_\alpha  \nonumber \\
\rho_{AB_2} & = \sum_\alpha p_\alpha \rho^A_\alpha \otimes \tau^{B_2}_\alpha
\end{align}
for some choice of states $\sigma^{B_1}_\alpha$ and $\sigma^{B_2}_\alpha$.  These notions are equivalent under channel-state duality.  Note that a single measure-and-prepare channel may sometimes be written using two different measurements, and likewise a single separable state may be written using two different ensembles.  

The main result of this paper is similar in spirit to a no-cloning or monogamy result, and likewise by the channel-state duality it will have nearly equivalent dynamical and static formulations, constraining either the dynamical properties of channels or the correlation properties of states.  

\section{Main result} \label{sec:main_result}

As discussed in Section \ref{sec:duality}, channel-state duality allows the result to be formulated as a statement about either channels or states.  We first describe Theorem \ref{thm:channels} for channels, because it is more directly related to the emergence of effective classicality described in Section \ref{sec:classicality}.  (The logic of the proofs, however, begins with Theorem \ref{thm:states} for states.)

Theorem \ref{thm:channels} considers arbitrary channels with many outputs, and it characterizes the reduced channels onto small subsets of outputs.  It states that for all small subsets of outputs except those overlapping some fixed $O(1)$-sized excluded subset, the corresponding reduced channels are measure-and-prepare, and moreover they use the same measurement. We denote this excluded region $Q$, or also the ``quantum Markov blanket.'' (The term ``Markov blanket'' follows terminology in classical statistics \cite{pearl1988probabilistic}.)

The result strengthens Theorem 2 of \cite{brandao2015generic}.

\begin{theorem} \label{thm:channels}
{\bf (Emergent classicality for channels.)}
Consider a quantum channel $\Lambda : \mathcal{D}(A) \to \mathcal{D}(B_1 \otimes ... \otimes B_n)$.  For output subsets $R \subset\{B_1,...,B_n\} $, let  $\Lambda_R \equiv \Tr_{\bar{R}} \circ \Lambda : \mathcal{D}(A) \to \mathcal{D}(R)$ denote the reduced channel onto $R$, obtained by tracing out the complement $\bar{R}$.  Then for any $\epsilon>0$ and $|R| \in \{1,...,n\}$, there exists a measurement, described by a positive-operator valued measure (POVM) $\{M_\alpha\}$, and an ``excluded'' output subset $Q \subset\{B_1,...,B_n\}$ of size
\begin{align} \label{eq:channel_thm_eqn}
|Q| \leq \frac{2 d_A^6 \ln(d_A)}{\epsilon^2} |R|
\end{align}
such that for all output subsets $R$ of size $|R|$ disjoint from $Q$, we have
\begin{align} \label{eq:channel_thm_eqn_eps}
\norm{\Lambda_R - \mathcal{E}_R}_\diamond \leq \epsilon
\end{align}
using a measure-and-prepare channel
\begin{align*} 
\mathcal{E}_{R}(X) := \sum_\alpha \Tr(M_\alpha X) \sigma^\alpha_R
\end{align*}
for some states $\{\sigma^\alpha_R\}_\alpha$ on $R$, where $d_A=\textrm{dim}(A)$ and $\norm{...}_\diamond$ is the diamond norm on channels.\footnote{The diamond norm on channels $N : \mathcal{D}(A) \to \mathcal{D}(B)$ is defined by \unexpanded{$\norm{N}_{\diamond} = \max_{\rho \in \mathcal{D}(AA')}  \norm{(N \otimes \mathds{1})(\rho_{AA'} )}_1$}.}  The measurement $\{M_\alpha\}$ does not depend on the choice of subset $R$, while the prepared states $\sigma^\alpha_R$ may depend on $R$.  
\end{theorem}

\begin{figure}
    \centering
    \includegraphics[width=3in]{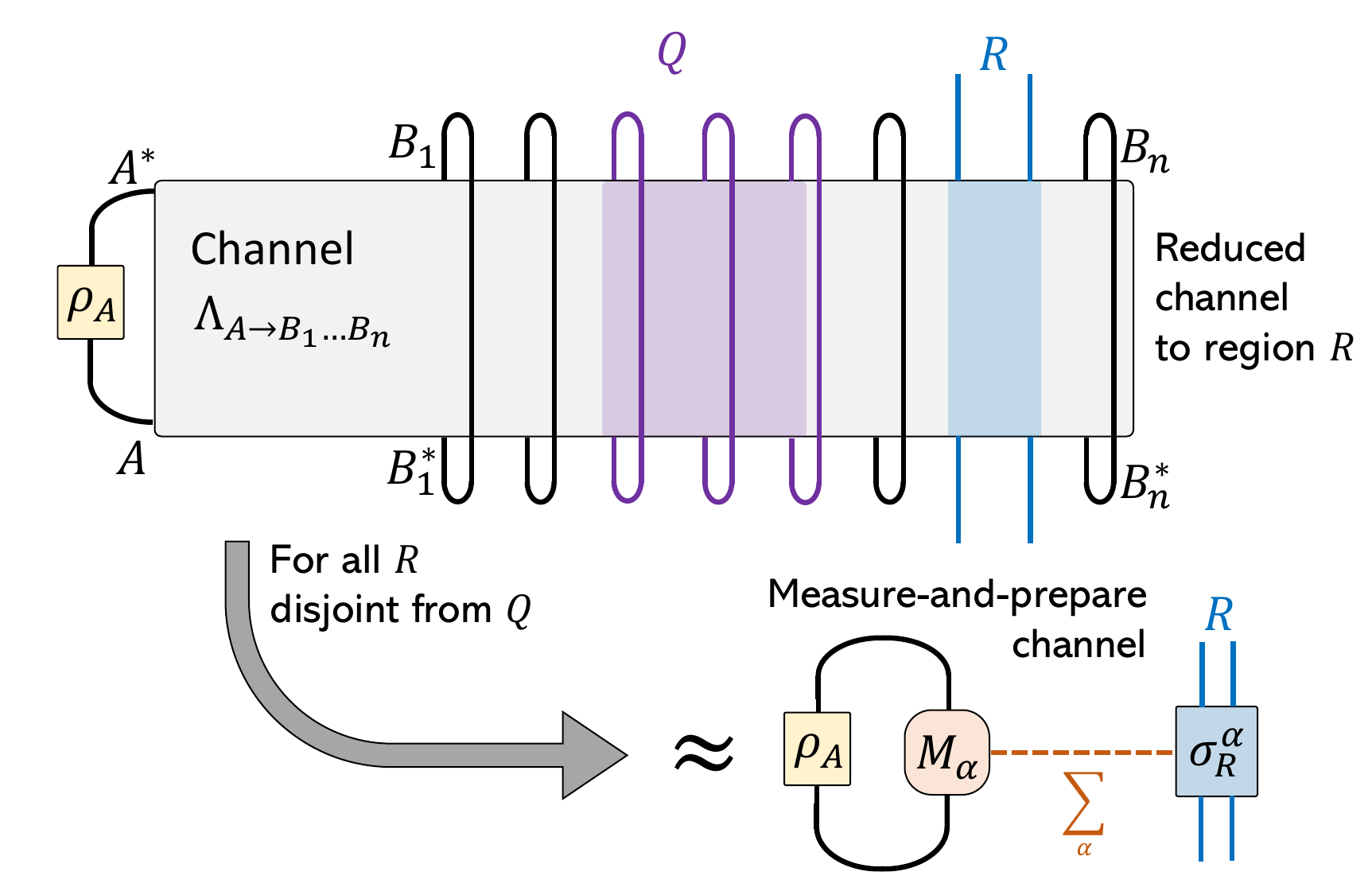}
    \caption{Illustration of Theorem \ref{thm:channels}. The quantum channel $\Lambda$ is shown acting on a state $\rho_A$, with a partial trace over the complement of the output region $R$. For any $R$ that does not overlap the ``Markov blanket'' $Q$, the reduced channel is approximately a ``measure-and-prepare'' channel. Importantly, the measurements $M_\alpha$ on $A$ are {\it independent} of the choice of region $R$.}
    \label{fig:channel}
\end{figure}

\noindent This theorem is illustrated in Fig.\ \ref{fig:channel}. 

\textbf{Variations of Theorem \ref{thm:channels}.} A stronger version of Eqns.\ \ref{eq:channel_thm_eqn}, \ref{eq:channel_thm_eqn_eps} in Theorem \ref{thm:channels} uses a variant of the diamond norm. This version is described in Eqn.\ \ref{eq:channel_theorem_1wayLOCC} and \ref{eq:diamondoneway}.  The following variation is weaker than Eqn.\ \ref{eq:channel_theorem_1wayLOCC} but perhaps simpler: one can replace Eqns.\ \ref{eq:channel_thm_eqn} and \ref{eq:channel_thm_eqn_eps} with
\begin{align} \label{eq:channel_theorem_max}
|Q|   \leq \frac{2 d_A \ln(d_A)}{\epsilon^2} |R|, \nonumber \\
 \max_{\rho \in \mathcal{D}(A)} \norm{\Lambda_R(\rho) - \mathcal{E}_R(\rho)}_1  \leq \epsilon.
\end{align}
Here and always in this work, $\norm{\cdot}_1$ denotes the Schatten 1-norm.  Note the superior dependence on $d_A$ above compared to that of Eqn.\ \ref{eq:channel_thm_eqn}, while constraining a weaker norm (when interpreting the entire LHS above as a norm).  We can also offer the following variation on Eqn.\ \ref{eq:channel_thm_eqn}: for any sizes $|R|$, $|Q|$, there exists region $Q$ of size $|Q|$ and POVM $\{M_\alpha\}$ such that for all regions $R$ of size $|R|$ disjoint from $Q$,
\begin{align} \label{eq:channel_thm_eqn_omega}
\norm{\Lambda_R - \mathcal{E}_R}_\diamond \leq d_A \Omega_{d_A, d_R} \sqrt{2 \ln{(d_A)}\frac{|R|}{|Q|}}
\end{align}
where $d_A = \dim(A), \, d_R = \dim(R)$ and $\Omega_{d_A, d_R}$ is the dimensional factor from Eqn.\ \ref{eq:omega_lemma},
\begin{align*}
\Omega_{d_A, d_R} = \min \{& d_A^2, 4d_A^{3/2}, 4d_R^{3/2}, 
& \sqrt{153 d_A d_R}, \, 2d_R-1\}. 
\end{align*}
Restricting the RHS above to $d_A^2$, we recover Eqn.\ \ref{eq:channel_thm_eqn}.  (For $d_R \gg d_A > 16$, the bound obtained by taking $\Omega_{d_A,d_R} = 4d_A^{3/2}$ is superior, whereas for $d_A \approx d_R$, the bound obtained from $\Omega_{d_A,d_R} = 2d_R-1$ is superior.)  The dimensional factor $\Omega_{d_A, d_R}$ arises from the relation between the ``one-way LOCC'' norm and the (Schatten) 1-norm; see more discussion surrounding Lemma \ref{lemma:onewaylocc}.  This dimensional factor may be sub-optimal and improved by future work.

As a final variation, in Eqn.\ \ref{eq:channel_thm_eqn_omega} and related expressions, we can also write a slightly tighter upper bound by using the replacement
\begin{align} \label{eq:replace_bound}
\sqrt{\frac{|R|}{|Q|}} \to \sqrt{\frac{1}{1+\left\lfloor \frac{|Q|}{|R|} \right\rfloor}}
\end{align}
where $\left\lfloor \cdot \right\rfloor$ is the integer floor function.  We use the simpler (but looser) bound only for readability; note we will need $|R|/|Q| \ll 1$ for the theorems to be useful, in which case the above replacement is only a small improvement.

\textbf{Remarks on Theorem \ref{thm:channels}.}
We refer to $Q$ as $O(1)$-sized because for a fixed size $|R|$ and error tolerance $\epsilon$, the size $|Q|$ depends on neither the total number of outputs nor the dimension of each output.  That is, the upper bounds in Eqn.\ \ref{eq:channel_thm_eqn} and elsewhere do not depend on $n$ or $\dim(B_i)$.  Physically, the region $Q$ is where any (non-negligible) locally accessible quantum information about $A$ must be stored. Therefore by no-cloning or monogamy of entanglement, no quantum information about $A$ can be locally accessible outside this region.  Meanwhile, $Q$ will also contain any locally accessible classical information about $A$.  However, unlike the quantum information, the classical information may also be present in copies outside of $Q$.

An essential point is that the measurement $\{M_\alpha\}$ in this theorem does not depend on $R$, so that apart from the $O(1)$-sized region $Q$,  different ``observers'' in different parts of the system can only receive classical information about the input in the same ``generalized basis,'' i.e.\ resulting from the same POVM on $A$.  (The observers may also receive no information at all.)  This supports the ``objectivity'' of the emergent classical description of quantum systems; see Section \ref{sec:discussion} for more discussion.

We now formulate the result for states rather than channels. 
\begin{theorem} \label{thm:states} 
{\bf (Emergent classicality for states.)}
Consider a quantum state $\rho \in \mathcal{D}(A \otimes B_1 \otimes ... \otimes B_n)$. Then for any $\epsilon>0$ and $|R| \in \{1,...,n\}$, there exist states $\{\rho_A^\alpha\}_\alpha$, probabilities $\{p_\alpha\}_\alpha$, and an ``excluded'' subset $Q \subset\{B_1,...,B_n\}$ of size 
\begin{align}
|Q| \leq \frac{2 \ln(2) S(A)}{\epsilon^2} |R|
\end{align}
such that for all subsets $R \subset\{B_1,...,B_n\}$ of size $|R|$ with $R\cap Q = \emptyset$, we have
\begin{align}
\norm{\rho_{AR} - \sum_\alpha p_\alpha \rho_A^\alpha \otimes \sigma_R^\alpha }_{\textrm{LOCC}_\leftarrow} \leq \epsilon,
\end{align}
for some choice of states $\{\sigma_R^\alpha\}_\alpha$ that depends on the choice of $R$.  The ensemble of states $\{p_\alpha, \rho^A_\alpha\}$ does not depend on the choice of $R$.  Here $S(A)$ denotes the von Neumann entropy of $\rho_A$, and the above ``one-way LOCC norm'' for bipartite states on $AR$ is defined as
\begin{align}
\norm{\rho_{AR}}_{\textrm{LOCC}_\leftarrow} \equiv \max_{M_R \in \textrm{QC}} \norm{(\mathds{1} \otimes M_R)(\rho_{AR})}_1,
\end{align}
with maximization taken over quantum-classical channels $M_R$ on $R$ (see Eqn.\ \ref{eq:QC} for a definition).
\end{theorem}

\textbf{Variations of Theorem \ref{thm:states}.} Note we can also replace $S(A)$ above with its upper bound $\log_2(d_A)$.

For two bipartite states $\rho, \sigma$ on $AR$, the above ``one-way LOCC norm'' \cite{matthews2009distinguishability, brandao2011faithful}, commonly denoted  $\norm{\rho-\sigma}_{\textrm{LOCC}_\leftarrow}$ or  $\norm{\rho-\sigma}_{\textrm{LOCC}_1}$, is related to the maximum probability of distinguishing between $\rho$ and $\sigma$ using local operations on $A$, $R$ and ``one-way'' classical communication from $R$ to $A$ (but not vice versa).\footnote{The arrow in $\norm{\cdot}_{\textrm{LOCC}_\leftarrow}$ indicates the direction of allowed communication.} It satisfies \begin{align}
    \norm{\rho_{AR}-\sigma_{AR}}_1 \leq  \Omega_{d_A,d_R} \norm{\rho_{AR}-\sigma_{AR}}_{\textrm{LOCC}_\leftarrow}
\end{align} 
with $\Omega_{d_A, d_R}$ as in Eqn.\ \ref{eq:omega_lemma},
\begin{align*}
\Omega_{d_A, d_R} = \min \{d_A^2, 4d_A^{3/2},4d_R^{3/2}, \sqrt{153 d_A d_R}, 2d_R-1\},
\end{align*}
with $d_A=\dim(A),\,d_R = \dim(R)$.  The above relation is introduced by Lemma \ref{lemma:onewaylocc} in Appendix \ref{sec:lemmas}.  In the context of Theorem \ref{thm:states}, we can then conclude 
\begin{align}
\norm{\rho_{AR} - \sum_\alpha p_\alpha \rho_A^\alpha \otimes \sigma_R^\alpha }_1 \leq \Omega_{d_A,d_R} \sqrt{2 \ln{(2)}S(A) \frac{|R|}{|Q|}}.
\end{align}
Finally, we also have the slight strengthening noted in Eqn.\ \ref{eq:replace_bound}.  

The state version of the theorem implies the channel version, by applying the state version to the Choi state of the channel.  Conversely, the channel version can only be used to directly prove the state version for states that have maximally mixed marginal on $A$, otherwise the state is not the Choi state of a channel.

\section{ Proofs} \label{sec:proof}

The proof builds on methods developed in \cite{brandao2017quantum, brandao2016product, brandao2015generic}.\footnote{The result in \cite{brandao2015generic} might initially appear to have superior dependence on $d_A = \dim(A)$ compared to Theorem \ref{thm:channels}, despite constraining fewer outputs.  But a side-by-side comparison reveals our Theorem \ref{thm:channels} actually has smaller error even for large $d_A$.  To make the comparison, first plug in $|Q| = \delta n$, where the LHS is in our notation and the RHS is in the notation of \cite{brandao2015generic}). Then note that for the bound in \cite{brandao2015generic} to be useful, one must also have $n \gg d_A^6 \log{(d_A)} $.}
First we will show Theorem \ref{thm:states} for states.  Afterward, we will use channel-state duality to obtain the theorem for channels.

We make use of the (quantum) mutual information, defined for a state $\rho$ on a system containing subsystems $X,Y$, as
\begin{align*}
I(X,Y)_\rho \equiv S(X)_{\rho} + S(Y)_{\rho} - S(XY)_{\rho},
\end{align*}
where $S(\cdot)$ is the von Neumann entropy.  We will suppress the state in the subscript when it is clear from context.  We also make use of the (quantum) conditional mutual information \cite{sutter2018approximate}, defined for a state $\rho$ on a system containing subsystems $X,Y,Z$,  as
\begin{align} \label{eq:CMI}
I(X,Y|Z)_\rho \equiv S(XZ) + S(YZ) - S(Z) - S(XYZ),
\end{align}
which one reads as ``the mutual information between $X$ and $Y$, conditioned on $Z$.''  The quantity is always non-negative, and the non-negativity is equivalent to strong subadditivity \cite{nielsen2004simple}.  Classically, the conditional mutual information quantifies how much information $X$ and $Y$ have about each other after conditioning on knowledge of $Z$. When the (quantum) conditional mutual information is small, the state on $XYZ$ forms an approximate (quantum) Markov chain \cite{fawzi2015quantum}.  In that case, the conditioned region $Z$ is sometimes referred to as a ``Markov blanket'' or ``Markov shield.''  The Markov blanket protects $X$ from direct correlations with $Y$ (or vice versa) in the sense that $X$ and $Y$ are independent when conditioned on the Markov blanket.  The region $Q$ of our main theorems is precisely such a Markov blanket. In other words, the correlations between $X$ and $Y$ are (almost) entirely mediated by their separate correlation with $Q$, if the conditional mutual information $I(X:Y|Q)$ (almost) vanishes.

The mutual information obeys a ``chain rule'' stating that for any state on subsystems $X,Y_1,...,Y_n$,
\begin{align} \label{eq:chain_rule}
I(X : Y_1 ... Y_n) = I(X:Y_1)  & + I(X : Y_2| Y_1)   \nonumber  \\ 
& + I(X : Y_3| Y_1 Y_2) \nonumber  \\ 
& +  ... + I(X : Y_n | Y_1...Y_{n-1}),
\end{align}
which can be verified by the definition of conditional mutual information, using a telescoping sum.  This simple equality may already be used to derive a monogamy result similar to Theorem \ref{thm:states} but not as powerful.  First note the LHS of Eqn.\ \ref{eq:chain_rule} is upper bounded by $2 \log(\dim(X))$, independent of $n$.  Because each of the $n$ terms on the RHS is positive and their sum has constant upper bound, most of them must be small.  In particular, for any $q$, no more than $q$ terms can be larger than $\frac{1}{q}$ times the upper bound.  So all but $q$ of the subsystems $Y_1,...,Y_n$ have the property that 
\begin{align}
I(X : Y_i | Y_1 ... Y_{i-1}) \leq 2 \log(\dim(X)) \frac{1}{q} .
\end{align}
When $X$ and $Y_i$ have low conditional mutual information conditioned on some third subsystem, they are close to separable.  (More precisely, the above LHS upper bounds the ``squashed entanglement'' $E_{sq}(X,Y_i)$ between $X$ and $Y_i$, which is an entanglement measure defined using conditional mutual information.  In \cite{brandao2011faithful} the authors demonstrate that states with small squashed entanglement are close to separable states, in the appropriate norm.)  So for most $Y_i$, the state on $X Y_i$ is close to separable.

The above statement is already close to the desired Theorem \ref{thm:states}, but it is weaker in an important way.  We want to prove not only that most reduced states on $X Y_i$ are close to separable, but also that they are close to separable when using a fixed ensemble of states on $X$ independent of $Y_i$, in the sense of Eqn.\ \ref{eq:same_ensemble}.  Equivalently, when we use channel-state duality to translate the claim to the channel setting, we want the reduced channels to be measure-and-prepare channels using the same measurement.  

The result we need is stated below in Proposition \ref{prop:one}, and it provides the core of the argument leading to Theorem \ref{thm:states} and then Theorem \ref{thm:channels}.  Proposition \ref{prop:one} also enables our improvement over the analogous results in \cite{brandao2015generic}, by using a simpler and more efficient optimization.

\begin{proposition} \label{prop:one}  \textbf{(Existence of small quantum Markov blankets.)}
Let $\rho_{AB_1...B_n}$ be a state on systems $A, B_1,...,B_n$, and choose any $|R|, q \in \{1,..,n\}.$  Then there exists a region $Q \subset\{B_1,...,B_n\}$ of size $|Q|\leq q$, along with quantum-classical channel (see Eqn.\ \ref{eq:QC}) $M_Q$ on $Q$, such that for all regions $R \subset\{B_1,...,B_n\}$ of size $|R|$ with $R \cap Q = \emptyset$,
\begin{align} \label{eqn:prop_one}
\max_{M_R \in \textrm{QC}} I(A : R | Q)_{M_Q M_R(\rho)} & \leq   S(A)_\rho \frac{1}{1+\left\lfloor \frac{q}{|R|}\right\rfloor} \\ 
& \leq  S(A)_\rho \frac{|R|}{q} , \nonumber
\end{align}
where $\left\lfloor \cdot \right\rfloor$ is the integer floor function, and the maximum is taken over all quantum-classical channels  $M_R$ on $R$.
\end{proposition} 
We refer to the region $Q$ as a \textbf{quantum Markov blanket} that ``covers'' or ``shields'' the region $A$.  See also the discussion below Eqn.\ \ref{eq:CMI}.  Proposition \ref{prop:one} essentially states that $A$ has small correlation with any region $R$ when conditioned on some measurement $M_Q$ on some sufficiently large region $Q$, and moreover this measurement need not depend on the choice of $R$.\footnote{The maximization over $M_R$ may be interpreted as qualifying the above statement to only refer to correlations between $A$ and $R$ that can still be discerned after performing a measurement on $R$.  Later, Lemma \ref{lemma:onewaylocc} will allow this maximization to be replaced with dimensional factors.}  The region $Q$ is ``small,'' or rather $O(1)$-sized, in the sense that for fixed error tolerance (when viewing the RHS of Eqn.\ \ref{eqn:prop_one} as an ``error,'' i.e.\ a deviation from zero conditional mutual information), $Q$ does not scale with the total number of subsystems $n$.

\begin{figure}
    \centering
    \includegraphics[width=3in]{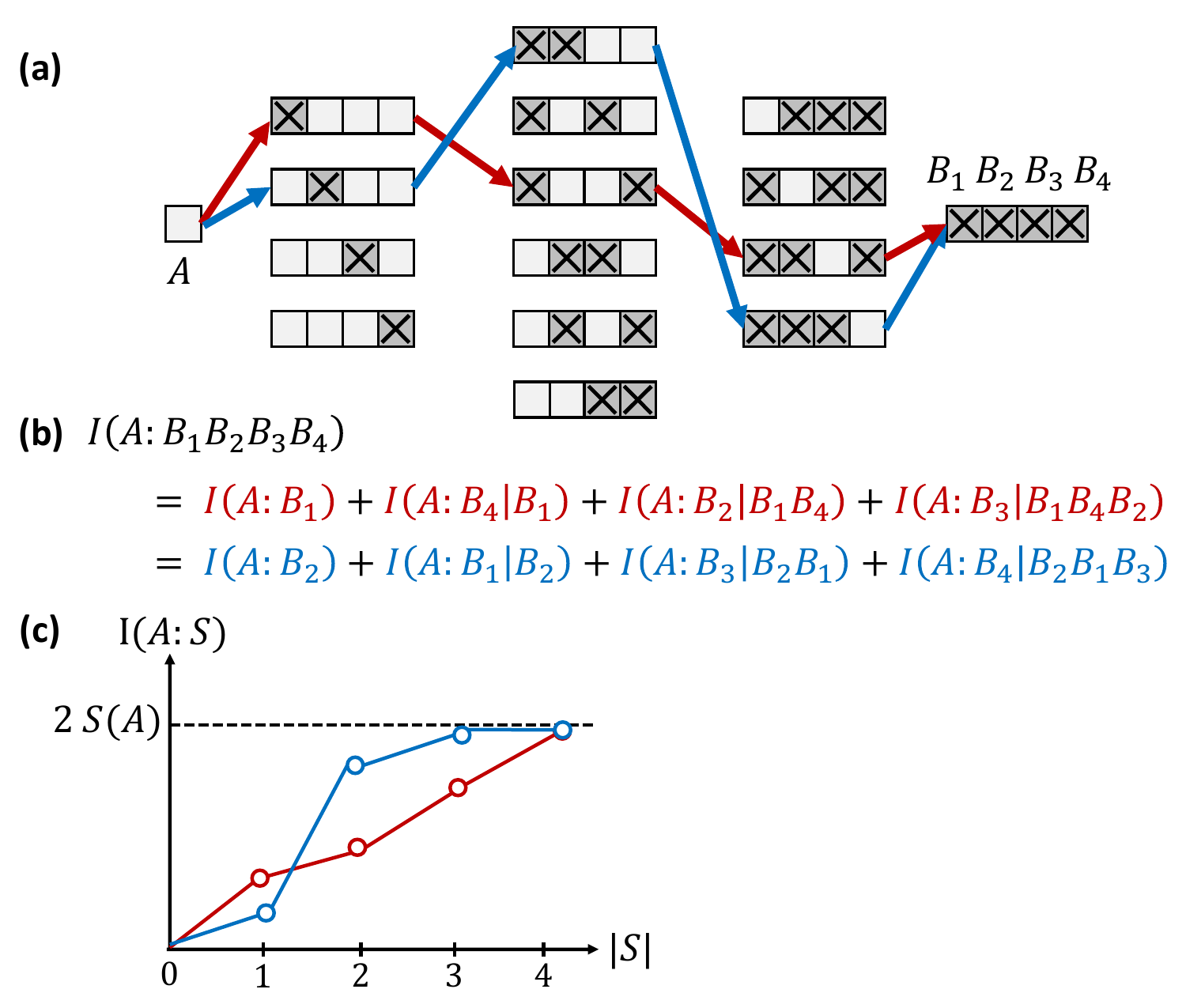}
    \caption{Illustration of the proof of Proposition \ref{prop:one}. 
     For simplicity we demonstrate the case of $n=4$ sites (qudits), with regions of size $|R|=1$. In panel (a), each node is a row of four boxes indicating a region $S$, i.e.\ a subset of outputs, with $x$'s indicating the elements of the subset.  Subsets are ordered left-to-right with increasing size. The arrows indicate inclusion, and each path indicates an expanding subset of outputs. Along each such path, the mutual information $I(A:S)$ increases monotonically and reaches the maximum when $S$ reaches the entire system (panel (c)). The increase of mutual information in each step is given by some conditional mutual information, shown in panel (b), where the first red term corresponds to the first red arrow, and so on.  These are positive by strong subadditivity. The proof of Proposition \ref{prop:one} considers the greedily optimized path, chosen by maximizing the terms in panel (b) from left to right.  Because the mutual information has a constant upper bound, for a long enough path we are guaranteed to find a ``bottleneck,'' where the conditional mutual information along any subsequent edge to any subsequent node must be small. Note that the mutual information is actually computed after applying a quantum-classical channel, which must also be optimized (as in the main text).}
    \label{fig:flow}
\end{figure}

\textbf{Proof of  Proposition \ref{prop:one}.}  A visual representation of the argument is sketched in Fig.\ \ref{fig:flow} for the case of $n=4$, $|R|=1$ and summarized in the caption. 

First, choose the region $S_1 \subset \{B_1,...,B_n\}$ of size $|R|$ and the quantum-classical channel $M_{S_1}$ on $S_1$ such that $S_1$ and $M_{S_1}$ together maximize $I(A,S_1)_{M_{S_1}(\rho)}$.
Next, choose the region $S_2 \subset \{B_1,...,B_n\}$ of size $|R|$, disjoint from $S_1$, and the quantum-classical channel $M_{S_2}$ on $S_2$ such that $S_2$ and $M_{S_2}$ together maximize the quantity
\begin{align*}
I(A,S_2|S_1)_{M_{S_2} M_{S_1}(\rho)}.
\end{align*}

Continuing, choose the region $S_3 \subset$~$\{B_1,...,B_n\}$ of size $|R|$, disjoint from $S_1 \cup S_2$, and the quantum-classical channel $M_{S_3}$ on $S_3$ so that $S_3$ and $M_{S_3}$ together maximize the quantity 
\begin{align*}
I(A,S_3|S_1 S_2)_{M_{S_3} M_{S_2} M_{S_1}(\rho)}.
\end{align*}

We continue choosing regions $S_i$ and quantum-classical channels $M_{S_i}$ in this way, until we have chosen $m$ regions $S_1,...,S_m$, where
\begin{align*}
    m = 1+\left\lfloor \frac{q}{|R|} \right\rfloor
\end{align*}
and $\left\lfloor \cdot \right\rfloor$ is the integer floor function.   We choose this number $m$ because ultimately the region $Q$ will be chosen as a subset of $S_1 \cup ... \cup S_{m-1}$, so that $Q$ will have size at most $(m-1)|R| \leq q$, as required.

By the chain rule of conditional mutual information (Eqn.\ \ref{eq:chain_rule}), we have
\begin{align} \label{eq:CMI_sum}
I(A,S_1...S_m)_{M(\rho)} & =  I(A,S_1)_{M(\rho)} + I(A,S_2|S_1)_{M(\rho)} \nonumber \\ 
& \;\; \; \; + ... + I(A,S_m|S_{m-1}...S_1)_{M(\rho)} \nonumber \\
& \leq  S(A)_\rho
\end{align}
for quantum-classical channel $M=M_{S_1}...M_{S_m}$ on $S_1...S_m$, where $S(A)_\rho$ is the von Neumann entropy of $A$ (not to be confused with the notation for regions $S_i$).  The latter inequality follows because $M(\rho)$ is separable between $A$ and $S_1...S_m$, and the mutual information of a bipartite separable state is at most the von Neumann entropy of either subsystem.  Note also $S(A) \leq  \log_2(d_A)$.

The LHS of the inequality in Eqn.\ \ref{eq:CMI_sum} has $m$ terms, each of which is positive by strong subadditivity.  Then the average term is at most $m^{-1} S(A)$, 
and at least one of the terms must be less than or equal to the average. Denote this term the $i^{th}$ term.  Then
\begin{align*}
I(A:S_i|S_1...S_{i-1})_{M_{S_1}...M_{S_i}(\rho)} \leq m^{-1} S(A)_\rho.
\end{align*}
Moreover, by our construction of $S_i$ and $M_{S_i}$, these choices maximized the LHS above.  So for any region $R$ of size $R$ disjoint from $S_1...S_{i-1}$, and for any quantum-classical channel $M_R$ on $R$,
\begin{align*}
I(A:R|S_1...S_{i-1})_{M_R M_{S_{i-1}}...M_{S_1}(\rho)} \leq m^{-1} S(A)_\rho.
\end{align*}
Letting $Q = S_1...S_{i-1}$, we have obtained the desired result, and $|Q| \leq |R|(m-1) \leq q$ by construction.
$\blacksquare$

\textbf{Proof of Theorem \ref{thm:states} for states.} The proof of Theorem \ref{thm:states} for states now proceeds as follows.  We begin with the setup and conclusion of Proposition \ref{prop:one}.  We conclude that for any $q$, there exists a region $Q \subset \{B_1,...,B_n\}$ of size $|Q| \leq q$, along with quantum-classical channel $M_Q$ on $Q$, such that for all regions $R  \subset \{B_1,...,B_n\}$ of size $|R|$ with $R \cap Q = \emptyset$, for all quantum-classical channels $M_R$ on $R$,
\begin{align*}
 I(A : R | Q)_{M_Q M_R(\rho)} \leq S(A)_\rho \frac{|R|}{q}.
\end{align*}
Then we apply Lemma \ref{lemma:closetoseparable} from Appendix \ref{sec:lemmas} to the state $M_Q M_R(\rho_{AQR})$ to conclude there exist probabilities $p_\alpha$ and states $\rho_A^\alpha, \rho_R^\alpha$ such that
\begin{align} \label{eq:thm_oneway}
\max_{M_R \in QC} \norm{(\mathds{1} \otimes M_R)\left(\rho_{AR} -  \sum_\alpha p_\alpha \rho_A^\alpha \rho_R^\alpha \right)}_1 \nonumber \\
\leq \sqrt{2 \ln{(2)}S(A)} \sqrt{\frac{|R|}{q}},
\end{align}
with maximum again over quantum-classical channels $M_R$ on $R$.  Note that the quantities $p_\alpha, \rho_A^\alpha$ produced by Lemma \ref{lemma:closetoseparable} depend only on $\rho_{AQ}$ and $M_Q$, not on the choice of $R$.

We have nearly arrived at the conclusion of Theorem \ref{thm:states}.  Note that if $|Q| < q$, we can add $q-|Q|$ arbitrary extra subsystems to $Q$ so that $|Q|=q$. Then using this enlarged region, Eqn.\ \ref{eq:thm_oneway} holds \textit{a fortiori} for all $R$ with $R \cap Q = \emptyset$, and for simplicity we formulate Theorem \ref{thm:states} without the $q$ parameter of Proposition \ref{prop:one}.
 
Thus we arrive at the conclusion of Theorem \ref{thm:states} for states. $\blacksquare$

\textbf{Proof of Theorem \ref{thm:channels} for channels.}  Finally we argue Theorem \ref{thm:channels} for channels.  
Given channel $\Lambda : \mathcal{D}(A) \to \mathcal{D}(B_1 \otimes ... \otimes B_n)$, consider the Choi state
\begin{align*}
\rho_{A'B_1...B_n} = \Lambda(|\Gamma\rangle \langle \Gamma|_{AA'})
\end{align*}
for a maximally entangled state $|\Gamma\rangle_{AA'}$ and reference system $A'$ isomorphic to $A$. Then apply Theorem \ref{thm:states} for states to this Choi state.  We obtain
\begin{align}
\norm{\rho_{A'R} - \sum_\alpha p_\alpha \rho_{A'}^\alpha \otimes \sigma_R^\alpha }_{\textrm{LOCC}_\leftarrow} \leq \sqrt{2 \ln{(d_A)}\frac{|R|}{|Q|}}.
\end{align}
We can relate the above one-way LOCC norm to the (Schatten) 1-norm using Lemma \ref{lemma:onewaylocc}, to obtain 
\begin{align} \label{eq:state_thm_proof_almost}
\norm{\rho_{A'R} - \sum_\alpha p_\alpha \rho_{A'}^\alpha \otimes \sigma_R^\alpha }_1 \leq \Omega_{d_A, d_R} \sqrt{2 \ln{(d_A)}\frac{|R|}{|Q|}}
\end{align}
where again $d_A = \dim(A), \, d_R = \dim(R)$ and $\Omega_{d_A, d_R}$ is the dimensional factor from Eqn.\ \ref{eq:omega_lemma},
\begin{align*}
\Omega_{d_A, d_R} = \min \{& d_A^2, 4d_A^{3/2}, 4d_R^{3/2}, \sqrt{153 d_A d_R}, \, 2d_R-1\}.
\end{align*}

Eqn.\ \ref{eq:state_thm_proof_almost} is almost the desired conclusion of Theorem \ref{thm:channels}; we just need to translate between Choi states and channels.

Recall from Section \ref{sec:duality} that reduced channels correspond to reduced states of the corresponding Choi state, and measure-and-prepared channels correspond to separable Choi states.  So the first term on the LHS above is the Choi state of the reduced channel $\Lambda_R$, and the second term on the LHS is the Choi state of some measure-and-prepare channel.  In particular, referring to Eqn.\ \ref{eq:measure_prepare_Choi}, the second term on the LHS of Eqn.\ \ref{eq:state_thm_proof_almost} is the Choi state of the corresponding measure-and-prepare channel 
\begin{align*} 
\mathcal{E}_{R}(X) \equiv \sum_\alpha \Tr(M_\alpha X) \rho^\alpha_R 
\end{align*}
with $M_\alpha = d_A p^\alpha (\rho_{A'}^\alpha)^T$. 

Now we just need to relate the (Schatten) 1-norm for Choi states to the diamond norm for the corresponding channels. For any channels $N_1, N_2$ on $A$ with corresponding Choi states $\rho^{N_1}, \rho^{N_2}$, a well-known lemma (see e.g.\ Lemma 6 of \cite{brandao2015generic}) gives the relation
\begin{align} \label{eq:diamondnormonenorm}
\norm{N_1-N_2}_\diamond \leq d_A \norm{\rho^{N_1} - \rho^{N_2}}_1.
\end{align}
Finally, Eqn.\ \ref{eq:channel_thm_eqn_omega} directly follows from Eqn.\ \ref{eq:state_thm_proof_almost} and the above translations between channels and Choi states.  Restricting to $\Omega_{d_A,d_B}=d_A^2$, the conclusion of Theorem \ref{thm:channels} follows as well. $\blacksquare$  

Note the additional factors of $d_A$ in Theorem \ref{thm:channels} compared to Theorem \ref{thm:states}.  One $d_A^2$ factor arose from the factor of $d_A$ in Eqn.\ \ref{eq:diamondnormonenorm}. The other factors arose from the $d_A^2$, or more generally the $\Omega$ factor in Eqn.\ \ref{eq:omega_lemma}, stemming from Lemma \ref{lemma:onewaylocc}. 

\textbf{Proof of further variations of Theorem \ref{thm:channels}.}  Alternatively, we can obtain a result for channels which avoids the factor of $d_A^2$ or $\Omega$ noted above by translating directly from Equation \ref{eq:thm_oneway}.  In that case, we can modify Theorem \ref{thm:channels} for channels to conclude
\begin{align} \label{eq:channel_theorem_1wayLOCC}
\norm{\Lambda_R - \mathcal{E}_R}_{\diamond\,  \textrm{LOCC}_\leftarrow} \leq d_A \sqrt{2 \ln{(d_A)}\frac{|R|}{|Q|}}
\end{align}
where we have defined a modified diamond norm, the ``diamond norm restricted to one-way LOCC,'' defined for a channel $N : \mathcal{D}(A) \to \mathcal{D}(B)$ as 
\begin{align} \label{eq:diamondoneway}
\norm{N}_{\diamond\,  \textrm{LOCC}_\leftarrow} = \max_{\rho \in \mathcal{D}(AA'),\, M_B \in QC} \norm{ (M_{B}N \otimes \mathds{1})(\rho_{AA'} )}_1
\end{align}
with the maximization taken over quantum-classical channels $M_{B}$ on $B$.  Note the advantage of this bound compared to the statement of Theorem \ref{thm:channels} using the diamond norm: here we have only $d_A$ on the RHS rather than $d_A^3$. 

To interpret this norm, note that for two channels $N_1, N_2$, the distance $\norm{N_1 - N_2}_{\diamond\,  \textrm{LOCC}_\leftarrow} $ measures the maximum distinguishability of $N_1, N_2$ when feeding them some state $\rho_{AA'}$ entangled with a reference system $A'$ and then using  one-way LOCC on $A'$ and $B$ to distinguish the two outputs, i.e.\ using only local operations on $A', B$ and one-way classical communication from $B$ to $A'$.  We then also have 
\begin{align*}
\norm{N}_{\diamond\,  \textrm{LOCC}_\leftarrow} \geq \max_{\rho \in \mathcal{D}(A)} \norm{N(\rho_{A} )}_1.
\end{align*}
Applied to Eqn.\ \ref{eq:channel_theorem_1wayLOCC}, the above yields Eqn.\ \ref{eq:channel_theorem_max} of Theorem \ref{thm:channels}. $\blacksquare$

In closing, we note that some more naive extensions of the proof methods in \cite{brandao2015generic} would fail here, as described in the footnote.\footnote{One might naively guess that Theorem 1 of \cite{brandao2015generic} could be used to prove our Theorem \ref{thm:channels} with the following trick. First apply the former theorem, which excludes some region $Q$ that grows with $n$.  Then because $Q$ is large for large $n$, focus on the reduced channel to $Q$ and apply the theorem to this channel alone.  Iterate the result in this fashion until the remaining region $Q$ is $O(1)$-sized.  However, this method suffers two flaws.  First, for a fixed error tolerance, more careful analysis reveals that the the final region $Q$ will still grow with $n$, albeit more slowly.  Second, each iteration of the theorem yields a new measurement for the measure-and-prepare channels, and these measurements will generally be different.} 

\section{Examples and numerics } \label{sec:examples}

Because Theorem \ref{thm:channels} applies to any channel, it will be helpful to consider a few very different cases.  Take $A$ to be a single qubit, and take $B$ to consist of $n$ qubits $B_1,...,B_n$.  We discuss several simple examples before turning to a detailed numerical example.

\begin{itemize}
\item Let $\Lambda : \mathcal{D}(A) \to \mathcal{D}(B)$ be the constant channel that takes every input to some constant state on $B$.  Then all the reduced channels are also constant, and moreover they are measure-and-prepare channels in a trivial sense: they can be expressed as any measurement on $A$ followed by a preparation of some constant state, independent of the outcome of the measurement.  Thus  Theorem \ref{thm:channels} easily holds, and in fact the approximation has zero error, and one could even take the excluded region $Q$ to be the empty set.  

\item Let $\Lambda$ be a Haar-random isometry.  Then for $A$ fixed and $n$ large, the reduced channels on small subsets will be again be approximately constant channels.  Thus the theorem applies as before.

\item Let $\Lambda$ faithfully transmit $A$ to some $B_i$, while preparing an arbitrary state on the remaining outputs.  Then the reduced channel $\Lambda_{B_i}$ is the identity channel, and the excluded region $Q$ must consist of at least $B_i$.  The remaining reduced channels are constant channels, and thus the error in Theorem \ref{thm:channels} is already zero for $|Q|=1$.  
 
\item Let $\Lambda$ be the isometry $|0\rangle_A \mapsto |0...0\rangle_B, \, |1\rangle_A \mapsto |1...1\rangle_B$.  Then every reduced channel $\Lambda_{B_i}$ is a measure-and-prepare channel, measuring in the $0/1$ basis and likewise preparing the $0/1$ state.  Thus the error in Theorem \ref{thm:channels} is already zero for empty $Q$.
\end{itemize}

A final example will be demonstrated numerically.  Consider a qubit $A$ that couples to a spin chain environment $E$ of $n-1$ qubits, $E = E_1 \otimes ... \otimes E_{n-1}$. The qubit begins in an arbitrary input state $\rho_A$, and the environment begins in some initial state $|\psi_0\rangle_E$.  Then the joint system $AE$ evolves unitarily under a joint Hamiltonian $H_{AE}$ for some time $t$.  Coupling the extra qubit to the spin chain produces the channel
\begin{align} \label{eq:numerics_example_channel}
\Lambda \, : \, & \mathcal{D}(A) \to \mathcal{D}(A \otimes E_1 \otimes ... \otimes E_{n-1}), \nonumber \\
& \rho_A \mapsto  e^{-iH_{AE}t} \left(\rho_A \otimes |\psi_0\rangle\langle\psi_0|_E \right) e^{i H_{AE}t}.
\end{align}
If desired, one may re-label the systems to obtain $\Lambda : \mathcal{D}(A) \to \mathcal{D}(B_1 \otimes ... \otimes B_n)$, matching the notation of Theorem \ref{thm:channels}.

For our numerical example, we take $H_{AE}$ to be the mixed-field Ising model with translation-invariant interaction term, with couplings chosen as in Eqn.\ 1 of \cite{banuls2011strong}, 
\begin{align*}
H = -\sum_{i=1}^{n-1} \sigma_z^i \sigma_z^{i+1} - g \sum_{i=1}^n \sigma_x^i - h \sum_{i=1}^n \sigma_z^i,
\end{align*}
in particular with $g=-1.05,\, h=0.5,$ so that the Hamiltonian is chaotic and far from any integrable model.  We take the initial environment state $|\psi_0\rangle_E$ to be the ground state of the same Hamiltonian restricted to $E$. We choose $H_{AE}$ to have open boundary conditions: we attach a single extra qubit $A$ to one end of an open spin chain with $n-1$ qubits.

Physically, we expect energy from $A$ to flow into the cool environment $E$, so this example is more representative of diffusion than of a measurement process.  However, it still illustrates the spread of information about $A$ into $E$.  

For short times, any information about $A$ will be confined to a small effective light-cone near the end of the chain where $A$ was attached.  The interior of this light-cone will constitute the optimized Markov blanket $Q$, and the reduced channels $A \to E_i$ for $E_i$ outside $Q$ will be nearly constant.  For longer times, the details depend on the dynamics of $H_{AE}$, and a larger $Q$ may be required to ensure the remaining reduced channels are close to measure-and-prepare.  However, for fixed error tolerance, Theorem \ref{thm:channels} guarantees $|Q|$ will have some finite maximum extent, independent of the size of $E$.  Thus $Q$ need not grow forever, even as the light-cone increases, and even if the environment were arbitrarily large.

\begin{figure}
    \centering
    \includegraphics[width=3in]{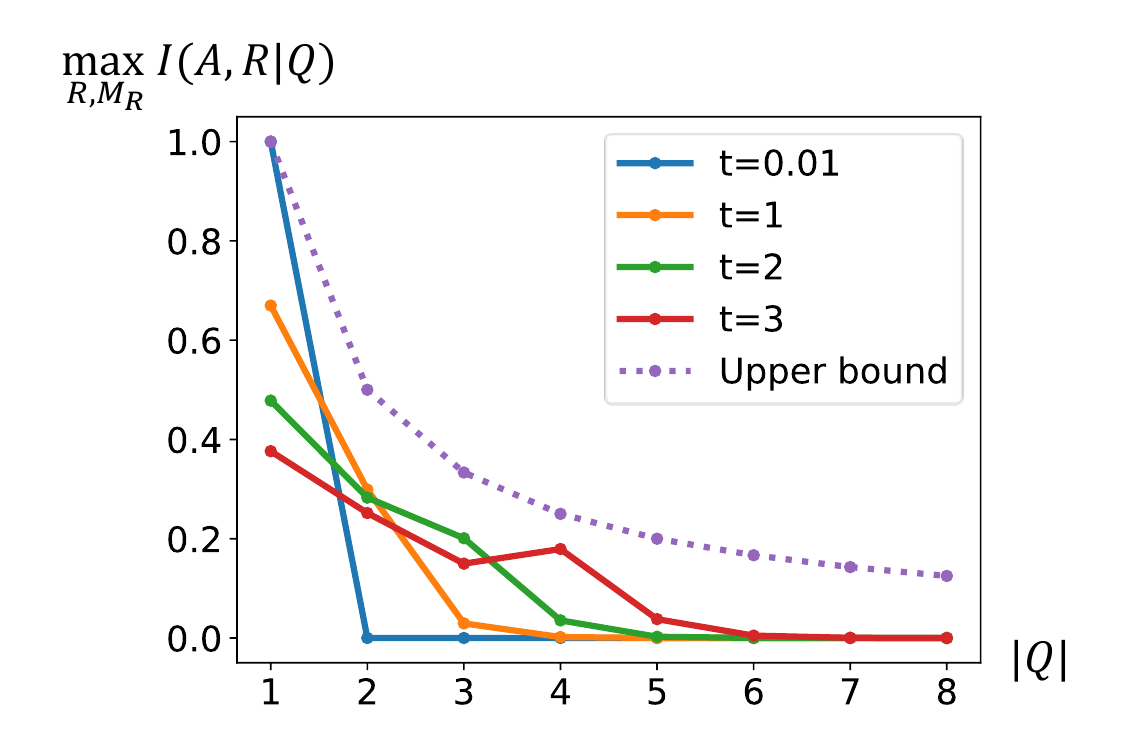}
    \caption{We consider an extra qubit $A$, in an arbitrary state $\rho_A$, attached to a spin chain of $7$ spins, which are initially in their ground state.  The qubit $A$ and the remaining $7$ spins initially have no correlation.  The $8$ spins then evolve for a time $t$ under a chaotic local spin chain Hamiltonian, giving rise to the channel in Eqn.\ \ref{eq:numerics_example_channel}.  For each $t$ and $|Q|=1,...,8$, we numerically calculate the optimal Markov blanket $Q$ of size $|Q|$, which best mediates the correlations between the input $A$ and the rest of the spin chain.  For the present example, in each case we find the optimal $Q$ consists of the $|Q|$ contiguous qubits at the end of the chain where $A$ was attached.  For the optimal $Q$, we plot the quantity $\alpha_Q$ of Eqn.\ \ref{eq:alpha_Q}, which has the interpretation of bounding the distance of the reduced channels (outside $Q$) to measure-and-prepare channels, as in Eqn.\ \ref{eq:alpha_Q_bound}.  We also plot the upper bound on $\alpha_Q$ given by Proposition \ref{prop:one}.  The figure demonstrates that at later times, a larger Markov blanket $Q$ is needed to ensure the remaining reduced channels are nearly  measure-and-prepare.  However, for fixed error tolerance, Theorem \ref{thm:channels} guarantees $|Q|$ to have some finite maximum extent. 
    }
    \label{fig:numerics}
\end{figure}

This example is depicted in Fig.\ \ref{fig:numerics}.   For each fixed $t$, and for each size $|Q|=1,...,n$, we construct an optimized Markov blanket $Q$ of size $|Q|$ and the associated optimal quantum-classical channel $M_Q$, for the case of $|R|=1$.  The procedure for constructing $Q$ is described in the proof of Proposition \ref{prop:one}.  

The construction involves an optimization over quantum-classical channels $M_R$ at each step.   Here, we further restrict to simple projective measurements with rank-1 projections.  Although this restricted optimization is not equivalent to an optimization over all quantum-classical channels, the result nonetheless implies the upper bounds of Theorem \ref{thm:channels}, because Eqn.\ \ref{eq:onewaylocc2} of Lemma \ref{lemma:onewaylocc} still holds for this restricted optimization.\footnote{Restricting the optimization to projective measurements with rank-1 projections is most efficient, but it does not entail the same optimum.  Alternatively, we can always restrict the optimization over quantum-classical channels, without loss, by restricting to quantum-classical channels with at most $\dim(R)^2$ outcomes and that use only rank-1 POVM elements.  To see this, first note that by convexity, an optimum will always occur on a so-called ``extremal POVM,'' and these have at most $\dim(R)^2$ outcomes \cite{d2005classical}.  Second, if any of the POVM elements of some optimum are not rank-1, the same optimum can be achieved with a rank-1 fine-graining, because the latter can be post-processed via coarse-graining into the original optimum, and the coarse-graining channel cannot decrease 1-norm.}
We perform the optimization numerically with a naive global optimization algorithm.

In Fig.\ \ref{fig:numerics}, for each $Q$, we plot the quantity
\begin{align} \label{eq:alpha_Q}
\alpha_Q \equiv \max_{R,M_R} I(A,R|Q)_{M_R M_Q(\rho)},
\end{align}
where the maximum is over all regions $R$ of size $|R|=1$ disjoint from $Q$, and all quantum-classical channels $M_R$ on $R$ (using only projective measurements).  The channel $M_Q$ is the optimal quantum-classical channel obtained together with $Q$.  The significance of the above quantity is that it upper bounds the distance of reduced channels $\Lambda_R$ to measure-and-prepare channels.  In particular, from the discussion around Eqn.\ \ref{eq:channel_theorem_max}, for all regions $R$ of the fixed size $|R|$ disjoint from $Q$, there is a measure-and-prepare channel $\mathcal{E}_R$ with measurement independent of $R$ such that
\begin{align} \label{eq:alpha_Q_bound}
\max_{\rho \in \mathcal{D}(A)} \norm{\Lambda_R(\rho) - \mathcal{E}_R(\rho)}_1 \leq d_A \sqrt{2 \ln{2} \alpha_Q}.
\end{align}

Fig.\ \ref{fig:numerics} also includes the upper bound on $\alpha_Q$ given by Proposition \ref{prop:one}.  Evidently it is not very tight, and so for this example the bound of Theorem \ref{thm:channels} is not tight either.  For other examples, it may be tighter.  Promising directions for future work include exploring the tightness of the bound, improving the bound for general states and channels, or alternatively improving the bound by specializing to a natural class of dynamics relevant to many-body physics.

\section{Further discussion } \label{sec:discussion}

We proved Theorem \ref{thm:channels} constraining the spread of quantum information in multi-output channels.  Alternatively, Theorem \ref{thm:states} constrains the correlation structure of multipartite states.  By constraining all but an $O(1)$ number of subsystems, these results give a much stronger constraint than the result of \cite{brandao2015generic}, which inspired the present work.  

To explicitly compare our Theorem \ref{thm:channels} with the analogous result of \cite{brandao2015generic}, observe that their Theorem 2 becomes
\begin{align} \label{eq:BPH_bound}
    |Q| \leq \frac{1}{\epsilon}3d_A^2\ln(d_A)^{\frac{1}{3}} n^{\frac{2}{3}} |R|^{\frac{1}{3}}
\end{align}
in our notation.  Compare the above to our Eqn.\ \ref{eq:channel_thm_eqn}, repeated for convenience,
\begin{align}
    |Q| \leq \frac{1}{\epsilon^2} 2 d_A^6 \ln(d_A)|R|. \nonumber
\end{align}
Their upper bound on $|Q|$ evidently scales with $n$, whereas our Eqn.\ \ref{eq:channel_thm_eqn} is independent of $n$.  While their bound may appear to have superior scaling with respect to $d_A$, $|R|$, and $\epsilon$, the bound is only useful when $n > |Q|$, in which case the RHS of Eqn.\ \ref{eq:BPH_bound} is at least 
\begin{align*}
    \frac{1}{\epsilon^3} 27 d_A^3 \ln(d_A) |R|,
\end{align*}
so in fact our bound is tighter in all regimes.

\subsection{Emergent classicality } \label{sec:classicality}

One significant motivation is to explain the emergence of the effective classicality of the quantum world, as discussed in \cite{brandao2015generic}.  An important ingredient in any such explanation is decoherence \cite{schlosshauer2019quantum}.  Suppose a previously isolated system $A$ interacts with a large environment $B$. Trace out $A$ and consider the resulting channel $A \to B$.  According to the standard narrative of decoherence, if the environment decohered the system, then any reduced channel $A \to B_i$ must be measure-and-prepare, with the measurement taken in the ``pointer'' basis for $A$, determined by the details of the decoherence process \cite{zurek2003decoherence}.

Perhaps surprisingly, our results (beginning with those of \cite{brandao2015generic}) demonstrate that an aspect of this classical structure exists in \textit{all} large states and channels.  Proceeding with the previous example, let us first examine a less interesting case. It is possible that after the interaction, $A$ is maximally entangled with $B_1$.  In that case, there is little sense in which $A$ has been robustly measured in some pointer basis: no systems other than $B_1$ have obtained any knowledge of $A$, so the information about $A$ has not spread. Regardless, Theorem \ref{thm:channels} holds.  The more interesting application of Theorem \ref{thm:channels} occurs when some information about $A$ \textit{does} become widely accessible to local observers $B_i$ in the environment.  In that case, Theorem \ref{thm:channels} states that the transmission of information $A \to B_i$ to these observers may be approximated as the result of some observer-independent measurement on $A$.  The POVM $\{M_\alpha\}$ produced by Theorem \ref{thm:channels} is effectively the pointer basis for this measurement process. 

In discussions of decoherence in many-body systems, often a particular subsystem is identified as ``the system,'' which is decohered by the remaining subsystems identified as ``the environment.''  This distinction may depend on particular features of the dynamics.  However, the authors of \cite{brandao2015generic} point out that their results (and by extension ours) remove the need for a presupposed split between system and environment; instead, we can choose any subsystem as the input system and treat the remaining subsystems as the environment.  Still, the decomposition of the total system into subsystems, including the decomposition of $B$ into regions $B_1,..., B_n$, may affect the POVM determined by Theorem \ref{thm:channels}, posing a question for future work.

The great generality of Theorems \ref{thm:channels} and \ref{thm:states} also leaves many important gaps in the explanation of emergent classicality.  On one hand, we have shown that for everywhere in the environment excluding an $O(1)$-sized region, any locally accessible information about a subsystem must be approximately classical.  On the other hand, as discussed in the examples of Section \ref{sec:examples}, in many cases the environment will contain \textit{no} locally accessible information about the subsystem.  In these cases, there may be no effective classical description of the dynamics.  For instance, the computational degrees of freedom inside a quantum computer certainly have no such description.  

Given that not all dynamics exhibit effective classicality, one must still ask what type of many-body dynamics allow a such an effective description, and which subsystems or degrees of freedom in particular exhibit this classicality. See \cite{riedel2010quantum, zwolak2014amplification, foti2019whenever, korbicz2017generic} for some discussion of this nature.

\subsection{ Compatible channels and states } \label{sec:compatibility}

Our results may also be framed in terms of the theory of compatibility \cite{coles2011information, heinosaari2016invitation}.  On a tripartite system $A B_1 B_2$, two reduced states (or ``marginals'') $\rho_{AB_1}$ and $\rho_{AB_2}$ are ``compatible'' if there exists a joint state $\rho_{AB_1 B_2}$ with those marginals.  Similarly, two channels $\Lambda_{B_1} : \mathcal{D}(A) \to \mathcal{D}(B_1)$ and $\Lambda_{B_2} : \mathcal{D}(A) \to \mathcal{D}(B_2)$ with the same input system are called ``compatible'' if there exists a joint channel $\Lambda : \mathcal{D}(A) \to \mathcal{D}(B_1 \otimes B_2)$ whose reduced channels are given by $\Tr_{B_2} \circ \Lambda = \Lambda_{B_1}$ and $\Tr_{B_2} \circ \Lambda = \Lambda_{B_2}$. Physically, channels are compatible when one can implement both channels simultaneously on the same input.   Reduced channels are compatible iff the corresponding Choi states are compatible, and the above discussion easily generalizes to larger multipartite systems.  No-cloning and monogamy of entanglement provide simple examples of compatibility constraints. No-cloning manifests as the incompatibility of any two unitary channels $A \to B_1$ and $A \to B_2$, while monogamy manifests as the incompatibility of any two maximally entangled states $\rho_{AB_1}$ and $\rho_{AB_2}$. 

Two measure-and-prepare channels that can be expressed using the same measurement are always compatible.  The converse is not true in general: there exist compatible measure-and-prepare channels that cannot be written using the same measurement (see Appendix \ref{sec:compatibility_example}).

From the perspective of compatibility, Theorem \ref{thm:channels} states that for any large collection of compatible channels, all but $O(1)$-many channels must be approximately measure-and-prepare, and moreover, they must be expressible using the same measurement. The existence of compatible channels that do not arise from the same measurement,  shown in Appendix \ref{sec:compatibility_example}, highlights the non-trivial nature of the second statement.

\subsection{ Previous monogamy-related results}

Quantum de Finetti theorems characterize the marginals of permutation-invariant states, which are approximately separable for large systems \cite{caves2002unknown}.  Thus de Finetti theorems corroborate the monogamy of entanglement.  Our result may be seen as a quantum de Finetti-type theorem for non-permutation-invariant systems. For instance, the result about $k$-extendible states in Corollary 2 of \cite{brandao2011faithful} may be seen as a special case of our Theorem \ref{thm:states} when specialized to permutation-invariant states.  Likewise, compare to Theorem 1 of \cite{brandao2017quantum}.

Early work in the direction of de Finetti-type results for non-permutation-invariant systems includes the ``decoupling'' theorems of \cite{brandao2016product}. These show that for large multipartite states, after conditioning the state on a measurement of a small random subset of qudits, the marginals on most other small subsets are approximately product states. (The measurement ``decouples'' them.)  The result of \cite{brandao2015generic} and our Theorem \ref{thm:states} may also be seen as decoupling theorems in this sense.  

The technique of using small conditional mutual information $I(X,Y|Z)$ to show $\rho_{XY}$ is close to separable was developed by \cite{brandao2011faithful}, where they use the one-way LOCC norm.  The use of the one-way LOCC norm in Theorem \ref{thm:states}, supported by Lemma \ref{lemma:onewaylocc}, is a technique inspired by \cite{brandao2017quantum}, where it was applied to obtain de Finetti theorems.  The method is further developed by \cite{brandao2017quantum, brandao2015generic, li2015quantum}.

In \cite{koashi2004monogamy} the authors demonstrate the the tradeoff between quantum and classical correlations.  In particular, if $A$ and $B$ have near-maximal classical correlation, then $A$ cannot have quantum correlations with any other system.   Using this result, one can show that in the setup of our Theorem \ref{thm:channels}, if even a single system $B_i$ receives near-maximal classical information about $A$, then automatically the other reduced channels must be approximately measure-and-prepare.  This fact also relates to the discussion about ``objectivity of outcomes'' in \cite{brandao2015generic}.  However, our results, and those of \cite{brandao2015generic}, do not require that any subsystem of the environment receives near-maximal classical information about $A$.  

\subsection{Future work}

There are many opportunities for future work.  The optimality of Theorems \ref{thm:channels} and \ref{thm:states} is unknown.  Certainly many channels will fail to saturate the inequalities.  Are the bounds tight for some channels, or can they be generally improved?  Some dependence on the dimension $d_A$ of the input system is necessary, but the exact dependence is unclear.   Likewise, the optimality of our bound with respect to the size $|Q|$ of the excluded region is also unclear.

None of our results depends on the size of the environmental subsystems $B$.  Moreover, Proposition \ref{prop:one} and Theorem \ref{thm:states} have no explicit dependence on the dimension $d_A$ of the input system.  These results already hold naively for infinite-dimensional inputs, as long as the state has finite entropy $S(A)$.  On the other hand, more care is required in Theorem \ref{thm:channels} for channels when $A$ is infinite-dimensional. 

References \cite{knott2018generic, colafranceschi2020refined} extend the results of \cite{brandao2015generic} to infinite-dimensional input systems $A$. Essentially, they replace the dimensional dependence with the assumption that the system $A$ has bounded energy.  The energy is taken with respect to some reference Hamiltonian; if the Hamiltonian's density of states does not grow too quickly, then the energy constraint implies an entropy constraint, which then replaces the dimensional dependence.  We imagine similar techniques could be used to extend our results to infinite-dimensional systems, combining our Proposition \ref{prop:one} with the tools developed in \cite{knott2018generic, colafranceschi2020refined}.

We are motivated by the emergence of effective classical descriptions of quantum many-body systems.  While our results demonstrate that some aspects of classicality are generic, an effective classical description requires more detailed properties of the dynamics.  Identifying these properties is an important area of research. Moreover, the bound in Theorem \ref{thm:channels} might be improved by specializing to some natural class of dynamics relevant for many-body physics.

Finally, this effective classicality suggests to us there exist efficient classical simulations of some quantum many-body systems.  We hope our numerical method in Section \ref{sec:examples} for determining the quantum Markov blanket and effective measurements may be useful here.

\begin{acknowledgments}
We wish to thank Patrick Hayden, Ludovico Lami, Jess Riedel, and Mark Wilde for valuable discussions. This work is supported by the National Science Foundation under grant $\#$1720504, and by the Simons Foundation. This work is also supported in part by the
DOE Office of Science, Office of High Energy Physics, the grant DE-SC0019380 (XLQ, DR).
\end{acknowledgments}

\appendix

\section{Lemmas}  \label{sec:lemmas}

The following lemma simply combines a collection of lemmas from \cite{brandao2015generic, aubrun2020universal, matthews2009distinguishability, lami2018ultimate}.

\begin{lemma} \label{lemma:onewaylocc} Let $L_{AB}$ be any Hermitian operator on $AB$.  Then
\begin{align}
\norm{L_{AB}}_1 \leq  \Omega_{d_A, d_B} 
\norm{L_{AB}}_{\textrm{LOCC}_\leftarrow}
\end{align}
where $d_A = \dim(A), \, d_B = \dim(B)$, and $\Omega_{d_A, d_B}$ is a dimensional factor
\begin{align} \label{eq:omega_lemma}
\Omega_{d_A, d_R} = \min \{& d_A^2, 4d_A^{3/2}, 4d_R^{3/2}, \nonumber \\ 
& \sqrt{153 d_A d_R}, \, 2d_R-1\}
\end{align}
and where $\norm{\cdot}_{\textrm{LOCC}_\leftarrow}$ is the ``one-way LOCC norm''
\begin{align*}
\norm{L_{AB}}_{\textrm{LOCC}_\leftarrow} \equiv \max_{M_B \in \textrm{QC}} \norm{(\mathds{1} \otimes M_B)(\rho_{AB})}_1,
\end{align*}
with maximization taken over quantum-classical channels $M_B$ on $B$ (see Eqn.\ \ref{eq:QC}).  See Eqn.\ \ref{eq:onewaylocc2} and the remark below it for a further variation.
\end{lemma}
 
\textbf{Proof and references.}  Different choices for the value of $\Omega_{d_A,d_R}$ on the RHS of Eqn.\ \ref{eq:omega_lemma} arise from different results in the literature.  The case of $\Omega = d_A^2$ was shown in Lemma 5 of \cite{brandao2015generic}. (This case has the most straightforward proof.) The case of $\Omega = 4\min\{d_A,d_B\}^{3/2}$ follows from Corollary 9 of \cite{aubrun2020universal}.  The use of this result was brought to our attention by Lemma A9 of \cite{colafranceschi2020refined}. The case of $\Omega = \sqrt{153 d_A d_B}$ follows from Theorem 15 of \cite{matthews2009distinguishability}; see also \cite{brandao2011faithful}.  Finally, the case of $\Omega = 2d_B-1$ follows from Theorem 16 of \cite{lami2018ultimate}.\footnote{The use of \cite{lami2018ultimate} for this purpose was brought to our attention by Ludovico Lami.}

We also note the following useful variant of Lemma 5 in \cite{brandao2015generic}, i.e.\ the case of $\Omega = d_A^2.$  There they showed
\begin{align} \label{eq:onewaylocc2}
\norm{\rho_{AB}}_1 \leq d_A^2 
\max_{M_B \in \textrm{QC}} \norm{(\mathds{1} \otimes M_B)(\rho_{AB})}_1,
\end{align}
as noted above, where the maximization is taken over quantum-classical channels $M_B$ on $B$.  \textit{Moreover, the inequality \ref{eq:onewaylocc2} still holds when the maximization is restricted to quantum-classical channels implemented by projective measurements, rather than more general POVMs.}

This slight strengthening of Lemma 5 of \cite{brandao2015generic} is useful for the numerical applications discussed in Section \ref{sec:examples}.  The proof of the modified lemma follows from the proof in \cite{brandao2015generic}  after noting that for any Hermitian operator $X$,
\begin{align*}
\norm{X}_1 = \max_{M} \norm{M(X)}_1
\end{align*}
where the optimization on the RHS yields the same answer whether taken over all channels $M$, just quantum-classical channels, or just quantum-classical channels implemented by projective measurements. $\blacksquare$

The next lemma we have excerpted from the proof in \cite{brandao2015generic}.

\begin{lemma} \label{lemma:closetoseparable} Adapted from the argument in \cite{brandao2015generic}.
Let $\rho_{ABC}$ be any state on $ABC$, let $M_C$ be any quantum-classical channel on $C$ (see Eqn.\ \ref{eq:QC}), and let
\begin{align*}
\epsilon = I(A:B|C)_{M_C(\rho)}.
\end{align*}
Then 
\begin{align*}
\norm{\rho_{AB} - \sum_\alpha p_\alpha \rho_A^\alpha \otimes \rho_B^\alpha}_1 \leq \sqrt{2 \ln{2} }\sqrt{\epsilon},
\end{align*}
where the quantum-classical channel $M_C$ measures POVM $\{M_C^\alpha\}_\alpha$ and
\begin{align*}
\rho_{AB}^\alpha & \equiv \frac{1}{p_\alpha} \Tr_C(\rho M_C^\alpha) \\
p_\alpha & \equiv \Tr(\rho M_C^\alpha).
\end{align*}
\end{lemma}

\noindent For convenience we repeat the argument used in \cite{brandao2015generic}.

\textbf{Proof.}  The state $M_C(\rho)$ is a quantum-classical state that is classical on $C$, i.e.\ 
\begin{align*}
\sum_\alpha p_\alpha \rho_{AB}^\alpha |\alpha\rangle \langle \alpha|_C
\end{align*}
with $p_\alpha$, $\rho_{AB}^\alpha$ as described in the lemma.  

Then direct calculation yields
\begin{align*}
\epsilon = I(A:B|C)_{M_C(\rho)} = \sum_\alpha p_\alpha I(A,B)_{\rho_{AB}^\alpha}.
\end{align*}

Note that in general 
\begin{align*}I(A,B)_\sigma = D(\sigma || \sigma_A \otimes \sigma_B) \geq \frac{1}{2 \ln{2}}\norm{\sigma - \sigma_A \otimes \sigma_B}_1^2,
\end{align*}
where $D(\cdot || \cdot)$ is the relative entropy and the inequality follows from quantum Pinsker's inequality.  Then
\begin{align*}
    \epsilon  & =  \sum_\alpha p_\alpha I(A,B)_{\rho_{AB}^\alpha} \\
    &  \geq \frac{1}{2 \ln{2}} \sum_\alpha p_\alpha \norm{\rho^\alpha_{AB} - \rho_A^\alpha \otimes \rho_B^\alpha}_1^2 \nonumber \\
    & \geq \frac{1}{2 \ln{2}}  \norm{\sum_\alpha p_\alpha( \rho^\alpha_{AB} - \rho_A^\alpha \otimes \rho_B^\alpha)}_1^2
\end{align*}
where the second inequality follows from the convexity of both the 1-norm and the function $x \mapsto x^2$.  The result follows. $\blacksquare$

\section{Compatible measure-and-prepare channels with distinct measurements} \label{sec:compatibility_example}

Measure-and-prepare channels are those which take the form (Eqn.\ \ref{eq:measure_and_prepare})
\begin{align*} 
\rho \mapsto \sum_\alpha \Tr(M_\alpha \rho)  \sigma_\alpha \nonumber
\end{align*}
for some POVM $\{M_\alpha\}$ and set of prepared states $\{\sigma_\alpha\}$.  Note that in general, this decomposition into a measurement and preparation is not unique; sometimes a different POVM and preparation yield the same channel.  

In this Appendix we demonstrate there exist measure-and-prepare channels that are compatible (in
the sense of \ref{sec:compatibility}) but that cannot be written using the same measurement.  That is, there exists some channel $\Lambda_{12} : \mathcal{D}(A) \to \mathcal{D}(B_1 \otimes B_2)$ for which the reduced channels 
\begin{align*}
    \Lambda_1 = \Tr_2 \cdot \Lambda_{12} : \mathcal{D}(A) \to \mathcal{D}(B_1) \nonumber \\
    \Lambda_2 = \Tr_1 \cdot \Lambda_{12} : \mathcal{D}(A) \to \mathcal{D}(B_2)
\end{align*} 
are both measure-and-prepare but cannot be expressed using the same POVM. 

For our example, take $A, B_1, B_2$ to be qubits, and define
\begin{align*}
\Lambda_1(\rho) & = \Tr(\rho|0\rangle\langle0|) \,|0\rangle\langle0| + \Tr(\rho |1\rangle\langle1|)\,|+\rangle\langle+| \nonumber \\
\Lambda_2(\rho) & = \Tr(\rho|+\rangle\langle+|) \, \rho_+ + \Tr(\rho |-\rangle\langle-|)\, \rho_-
\end{align*}
where 
\begin{align*}
|\pm \rangle & = \frac{1}{\sqrt{2}}\left(|0\rangle \pm |1\rangle\right) \nonumber \\
\rho_{+} & = p|0\rangle\langle0| + (1-p)|1\rangle\langle1| \nonumber \\
\rho_{-} & = (1-p)|0\rangle\langle0| + p|1\rangle\langle1|  
\end{align*}
for some $p\in[0,1]$. 
Then $\Lambda_1$ measures in the $|0\rangle,|1\rangle$ basis and prepares the non-orthogonal states $|0\rangle, |+\rangle$ contingent on the outcome.  On the other hand, $\Lambda_2$ measures in the $|+\rangle,|-\rangle$ basis and prepares the non-orthogonal states $\rho_+$, $\rho_-$ contingent on the outcome.

We want to demonstrate (a) $\Lambda_1, \Lambda_2$ are compatible, and (b) they cannot be re-expressed as measure-and-prepare channels using the same measurement.  

To show compatibility, we need to find a channel  $\Lambda_{12}$  with reduced channels $\Lambda_1, \Lambda_2$.  That amounts to finding a linear superoperator $\Lambda_{12}$ subject to the linear constraints $\Lambda_1 = \Tr_2 \cdot \Lambda_{12}$, $\Lambda_2 = \Tr_1 \cdot \Lambda_{12}$ as well as the inequality that $\Lambda_{12}$ is completely positive, or equivalently that the Choi state is a positive operator.  One solution is the channel with Choi state $\rho_{AB_1B_2}$, where the reference system of the Choi state is identified with $A$,
\begin{align*}
\rho_{AB_1B_2} = &\frac{1}{4}(|000\rangle\langle000| + |001\rangle\langle001|) \nonumber \\
& + \frac{1}{4}(|1+0\rangle\langle1+0| + |1+1\rangle\langle1+1|) \nonumber \\
& + \frac{\sqrt{2}}{2}(p-\frac{1}{2})(|000\rangle\langle1+0| - |001\rangle\langle1+1|) \nonumber \\
& + \textrm{h.c.}
\end{align*}
One can verify by inspection that the reduced states $\rho_{AB}$ and $\rho_{AC}$ coincide with the Choi states of the reduced channels $\Lambda_1, \Lambda_2$.  To verify $\Lambda_{12}$ is a valid channel, we need only verify it is completely positive, or equivalently that $\rho_{AB_1B_2}$ is a positive operator.  Diagonalizing the above $8$-by-$8$ matrix, one finds the eigenvalues are positive for $p \in [\frac{1}{2}-\frac{1}{2\sqrt{2}},\frac{1}{2}+\frac{1}{2\sqrt{2}}]$.  Thus for any such $p$, the channels $\Lambda_1, \Lambda_2$ are compatible.

To argue $\Lambda_1, \Lambda_2$ cannot be re-expressed using the same measurement,\footnote{We thank Patrick Hayden for useful comments leading to this argument.} first we show that any measure-and-prepare decomposition of $\Lambda_{1}$ must have its measurement in the $|0\rangle,|1\rangle$ basis.  Assume it could be written using a general POVM $\{M_\alpha\}$ and preparation of states $\{\sigma_\alpha\}$.   Then $\Lambda_1(|0\rangle\langle0|) =|0\rangle\langle0| = \sum_\alpha \Tr(|0\rangle\langle0|M_\alpha)\sigma_\alpha$.  But recall that pure states are extremal in the sense of convex sets, meaning that in general, if a pure state $|\psi\rangle\langle\psi|$ can be expressed as a positive sum of positive states $|\psi\rangle\langle\psi| = \sum_\alpha p_\alpha \rho_\alpha$, with $p_\alpha>0$, then $\rho_\alpha = |\psi\rangle\langle\psi|$ for all $\alpha$.  So in our case, for any $M_\alpha$ that overlaps $|0\rangle\langle0|$ (i.e.\ $\Tr(|0\rangle\langle0|M_\alpha)>0$), we must have $\sigma_\alpha=|0\rangle\langle0|$.  Likewise, because $\Lambda_1(|1\rangle\langle1|) =|+\rangle\langle+|$ and $|+\rangle\langle+|$ is also pure, we must have $\sigma_\alpha=|+\rangle\langle+|$ for any $M_\alpha$ that overlaps $|1\rangle\langle1|$.   Each $M_\alpha$ must overlap at least $|0\rangle\langle0|$ or $|1\rangle\langle1|$, and none can overlap both (which would require $\sigma_\alpha = |0\rangle\langle0|$ and also $\sigma_\alpha = |+\rangle\langle+|$), so we must have that each $M_\alpha$ is proportional to either $|0\rangle\langle0|$ or $|1\rangle\langle1|$, and they can be collected into two POVM elements $|0\rangle\langle0|$ and $|1\rangle\langle1|$, thus $\Lambda_1$ must measure in the $|0\rangle,|1\rangle$ basis as claimed.   

To finish we must argue $\Lambda_2$ cannot be expressed using a measurement in the $|0\rangle,|1\rangle$ basis.  Assume to the contrary $\Lambda_2(\rho) =  \Tr(\rho|0\rangle\langle0|) \, \sigma_0 + \Tr(\rho |1\rangle\langle1|)\, \sigma_1 $ for some states $\sigma_0, \sigma_1$.  Then direct calculation yields $\Lambda_2(|+\rangle\langle+|) =  \Lambda_2(|-\rangle\langle-|) = \frac{1}{2}(\sigma_1 + \sigma_0)$, so we would require $\rho_+ = \rho_-$, i.e.\ $p=\frac{1}{2}$.  Thus for $p \neq \frac{1}{2}$, $\Lambda_2$ cannot be expressed using the same measurement as $\Lambda_1$, as desired. $\blacksquare$

\bibliographystyle{unsrtnat}
\bibliography{references}

\end{document}